\documentclass[10pt,twocolumn,letterpaper]{article}

\usepackage{iccv}
\usepackage{times}
\usepackage{epsfig}
\usepackage{graphicx}
\usepackage{amsmath}
\usepackage{amssymb}
\usepackage[table,xcdraw]{xcolor}
\usepackage{cite}
\usepackage{multirow}

\usepackage[breaklinks=true,bookmarks=false]{hyperref}

 \iccvfinalcopy % *** Uncomment this line for the final submission

\begin{document}

%%%%%%%%% TITLE
\title{Adversarial Heart Attack: \protect\\
Neural Networks Fooled to Segment Heart Symbols in Chest X-Ray Images}
\author{Anonymous\\
Institution\\

}
\author{
Gerda Bortsova$^{1*}$, Florian Dubost$^{1}$\thanks{ indicates equal contribution.}, Laurens Hogeweg$^{2}$, Ioannis Katramados$^{2}$, Marleen de Bruijne$^{1,3}$\\
${^1}$Erasmus University Medical Center, Rotterdam, NL\\
${^2}$Intel Corporation, Santa Clara, USA\\
${^3}$University of Copenhagen, Copenhagen, DK
}

\maketitle
\begin{abstract}
Adversarial attacks consist in maliciously changing the input data to mislead the predictions of automated decision systems and are potentially a serious threat for automated medical image analysis. Previous studies have shown that it is possible to adversarially manipulate automated segmentations produced by neural networks in a targeted manner in the white-box attack setting (assuming the full access to the target model). In this article, we studied the effectiveness of adversarial attacks in targeted modification of segmentations of anatomical structures in chest X-rays: the heart, lungs, and clavicles. We focused on two following aspects not explored in previous studies. Firstly, we experimented with using anatomically implausible shapes as targets for adversarial manipulation. We showed that, by adding almost imperceptible noise to the image, we can reliably force state-of-the-art neural networks to segment the heart as a heart symbol instead of its real anatomical shape. Moreover, such heart-shaping attack did not appear to require higher adversarial noise level than an untargeted attack based the same attack method. Secondly, we attempted to explore the limits of adversarial manipulation of segmentations. For that, we assessed the effectiveness of shrinking and enlarging segmentation contours for the three anatomical structures. We observed that adversarially extending segmentations of structures into regions with intensity and texture uncharacteristic for them presented a challenge to our attacks, as well as, in some cases, changing segmentations in ways that conflict with class adjacency priors learned by the target network. Additionally, we evaluated performances of the untargeted attacks and targeted heart attacks in the black-box attack scenario, using a surrogate network trained on a different subset of images. In both cases, the attacks were substantially less effective. We believe these findings bring novel insights into the current capabilities and limits of adversarial attacks for semantic segmentation.

\end{abstract}

\section{Introduction}

\begin{figure*}[!t]
\centering
\includegraphics[width=\textwidth]{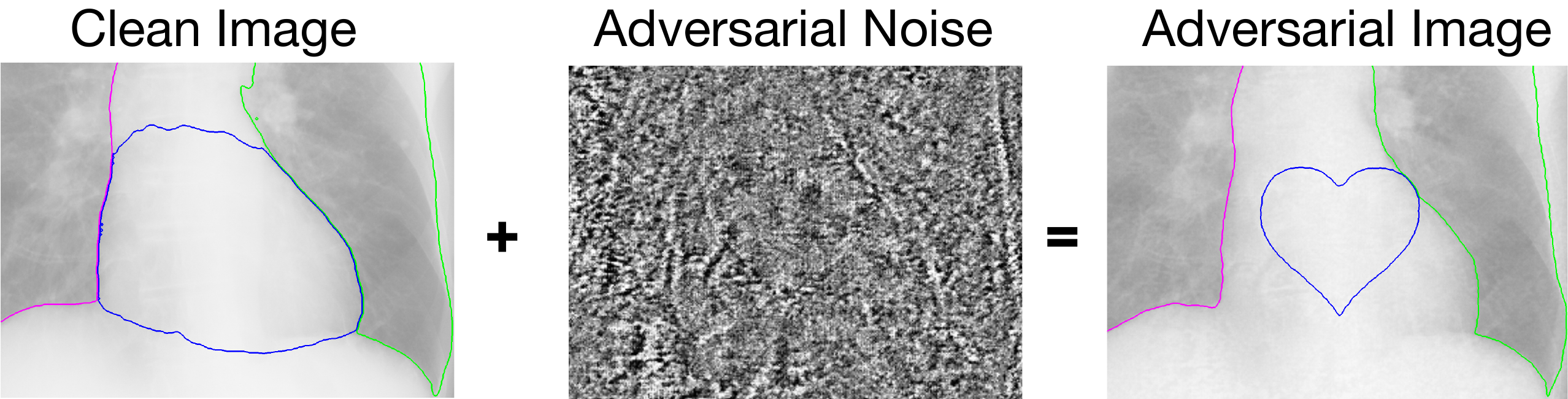}
\caption{\textbf{State-of-the-art segmentation network fooled to segment a heart symbol in chest X-Ray images.}
The original clean image with corresponding heart segmentation contours predicted by the network (blue contours) is shown on the left. We used a white-box projected gradient descent (PGD) \cite{madry2017} to generate adversarial noise shown in the middle. The last image shows the image with added adversarial noise and corresponding predicted segmentations.
The network was fooled by the noise and segmented the heart as a heart symbol instead of segmenting its real anatomical shape. This image is a representative sample: we repeated the same experiment in 20 random images and obtained an average intersection over union between the forced and predicted heart shape of 97.8\%.
}
\label{fig:heart}
\end{figure*} %

In the last years, convolutional neural networks (CNN) have achieved tremendous success in the field of medical image analysis. Simultaneously, CNNs have also been criticized to lack transparency in the computation of their predictions. Vulnerability of neural networks to \textit{adversarial attacks}, which manipulate the input to these algorithms to mislead their predictions, contributes to this lack of transparency \cite{szegedy2013intriguing,goodfellow2014,wang2018,ilyas2019,ford2019,corneanu2019,bubeck2019,akhtar2018,arnab2018,biggio2018}. Researchers have categorized adversarial attacks and corresponding defenses \cite{yuan2019, akhtar2018,biggio2018}. Many attacks consist in adding carefully computed noise to the input image under the assumption that this noise is imperceptible for humans. 

Adversarial attacks may have devastating consequences in healthcare by allowing unscrupulous organizations and individuals to manipulate the diagnosis computed by neural networks. For example, a healthcare provider could manipulate the patient diagnosis to seek higher insurance coverage \cite{finlayson2019}. To prevent such misuse, the potential and limits of adversarial attacks need to be better understood. 

In medical image analysis, adversarial attacks are gaining interest \cite{finlayson2019,taghanaki2018,paschali2018generalizability,ozbulak2019,ma2020}. Taghanaki et al. \cite{taghanaki2018} studied adversarial attacks for classification in chest X-ray, Paschali et al. \cite{paschali2018generalizability} in dermoscopy images, and Ma et al. \cite{ma2020} in fundoscopy, chest X-ray, and dermoscopy. Semantic segmentation, the most common task in medical image analysis, has been less researched. Xie et al. \cite{xie2017}, Hendrik et a. \cite{hendrik2017}, Fischer et al. \cite{fischer2017}, and Arnab et al. \cite{arnab2018robustness} have investigated adversarial attacks for segmentation networks in computer vision datasets.
They have shown that segmentation networks can be forced to output a specified target segmentation in the white-box setting, i.e. assuming the target network, including its weights, is fully available to the attacker \cite{xie2017,hendrik2017,fischer2017}.
Arnab et al. \cite{arnab2018robustness} showed that the segmentation performance of the target network can be reduced even when adversarial examples are crafted by a different network--called \textit{surrogate} model--which corresponds to the black-box attack setting. Paschali et al. \cite{paschali2018generalizability} studied adversarial attacks on brain segmentation networks and also showed that they can reduce segmentation accuracy in the black-box setting.
Xie et al. \cite{xie2017} demonstrated that a targeted attack, forcing predicted segmentations to a specific, erroneous target segmentation  was possible to some degree. Ozbulak et al. \cite{ozbulak2019} recently investigated attacks for segmentation networks in glaucoma optic disc segmentation. They showed that adversarial attacks could, in the white-box setting, make networks predict a target segmentation of the glaucoma optic disc that was anatomically plausible.

In this paper, we demonstrate that networks can also be fooled to predict anatomically implausible segmentations, which were never seen in the training set. As an example, we created adversarial examples that make the networks segment the heart as a heart symbols in chest X-rays, while keeping segmentations of other structures, namely, clavicles and lungs, accurate (Figure \ref{fig:heart}). We evaluated these heart-shaping attacks for heart symbols of different sizes. To further assess the limits of adversarially changing segmentations, we performed additional experiments in which the size of different structures was progressively increased or decreased compared to that of the ground truth segmentation. We investigated how the maximum magnitude of the adversarial noise affects the results, and whether similar performance can be achieved using black-box attacks instead of white box-attacks.

\section{Methods and materials}
\label{sec:methods}
In this section, we first describe the utilized adversarial attacks. Second, we explain the difference between white-box and black-box attacks. Third, we detail the chest X-ray dataset used for the experiments. Last, we describe the networks used in our experiments.

\subsection{Adversarial Attacks}

The principle of many adversarial attack methods is to modify an input image $x$ just slightly to obtain an adversarial example $x'$ that preserves the semantic content and thus the ``ground truth'' class $y$ of the original image $x$ but changes the prediction of the network with parameters $\theta$. 

Goodfellow et al. \cite{goodfellow2014} proposed the \textit{fast gradient sign method}, which computes an adversarial example as

\begin{equation}
\label{eq:fgsm}
    x' = x + \epsilon \cdot \text{sign}(\nabla_x J(\theta,x,y)),
\end{equation}
where $\epsilon$ is the maximum noise per pixel and $J$ is the loss function (e.g., the loss the attacked network was trained with). For image classification networks, the ground truth $y$ is of spatial dimension one. In this article, we studied segmentation networks, for which $y$ is a ground truth segmentation of the size of the input image. 

\textit{Projected gradient descent} (PGD) is a method proposed by Madry et al. \cite{madry2017} to compute stronger adversarial attacks. PGD is an iterative version of the fast gradient sign method that computes adversarial examples as

\begin{equation}
\label{eq:pgd}
x^{t+1} = \text{clip}_x^\epsilon \left( x^{t} + \alpha \cdot \text{sign}(\nabla_x J(\theta,x,y)|_{x = x^{t}}) \right),
\end{equation}
where $\text{clip}$ is a function that ensures that the magnitude of the added noise remains under a fixed limit $\epsilon$ around $x$, $x^0 = x$, and $\alpha$ is the step size. Computing $x^1$ with $\alpha = \epsilon$ is equivalent to an FGSM attack. We call $\epsilon$ the \textit{noise level} in the rest of the article.

Equations \ref{eq:fgsm} and \ref{eq:pgd} correspond to untargeted versions of FGSM and PGD, respectively: attacks that minimize the similarity between the predictions of the network and the ground truth, as measured by loss function $J$. A targeted PGD attack, maximizing the similarity between network predictions and a target segmentation $\tilde{y}$, can be computed as follows:
\begin{equation}
\label{eq:pgd}
x^{t+1} = \text{clip}_x^\epsilon \left( x^{t} - \alpha \cdot \text{sign}(\nabla_x J(\theta,x,\tilde{y})|_{x = x^{t}}) \right),
\end{equation}
As in the untargeted attack case, targeted FGSM attack can be obtained by setting the number of iterations to one and $\alpha = \epsilon$.

\subsection{White box versus black box attacks}
To create adversarial samples, the attacker needs to compute the gradients of the loss $\nabla_{x}J(\theta,x,\cdot)$, and needs therefore to have access to the target (i.e. attacked) network's architecture and weights $\theta$. However, accessing those may not be realistic for closed-source deep learning systems.
%Keeping the code secret might result from practical or commercial reasons but could also be employed as a defensive measure against adversarial attacks.
Attacks that require full access to the network's architecture and weights are called white-box attacks \cite{goodfellow2014}. Black-box attacks, which do not require the network's weights or architecture, are also possible. For example, another network, often called ``surrogate'' network, can be used instead of the target network to create adversarial examples \cite{papernot2017}.

\subsection{Dataset}
For the experiments presented in this article, we used the Japanese Society of Radiological Technology (JSRT) dataset \cite{shiraishi2000}. This dataset contains 247 posterior-anterior chest radiographs with a resolution $2048 \times 2048$, 0.175 mm pixel size and 12-bit depth. Ginneken et al. \cite{ginneken2006} later released segmentations for left and right lung fields, left and right clavicles, and the heart for a subset of images. To accelerate the computations, the images were resampled to a resolution of $512 \times 512$ pixels.

\subsection{Segmentation networks}
For both target and surrogate networks, we used a state-of-the-art convolutional neural network proposed by Bortsova et al. \cite{bortsova2019}. The architecture of this network was similar to that of a U-Net segmentation network \cite{ronneberger2015}. The differences compared to the original U-Net were as follows. The number of features in every layer was reduced by 4; strided convolutions were used instead of max pooling; batch normalization layers were used after every convolutional layer.
The resulting receptive field of this network is $184 \times 184$ pixels, i.e. it spans around one third of the image's width and hight.
The network was trained using the negative intersection over union (IoU) averaged over the six classes (the five anatomical structures and the background) as the loss function. We used Adadelta optimizer \cite{zeiler2012} with the default starting learning rate. As a data augmentation strategy, elastic deformations were applied to images half of the time. The deformation fields for elastic deformations were created by randomly sampling two-dimensional displacement maps from a uniform distribution U(-1000, 1000) and smoothing them with a Gaussian filter with the standard deviation of 100 pixels. Spline interpolation was applied to images and nearest neighbor interpolation was applied to labels and predictions.

The target and surrogate networks were trained using training sets of 100 images and validation sets of 23 images, obtained by randomly splitting the official training subset of the dataset.
Validation sets for the two networks were sampled such that they do not overlap.
The testing set consisted of 20 images randomly sampled from the official test subset.

\section{Experiments and Results}
We present the results of three series of experiments. First, we assessed the performance of untargeted attacks in the white-box and black-box settings with five levels of adversarial noise. Secondly, we 
assessed the performance of adversarially reshaping heart
segmentations into heart shapes with the same noise levels in the white-box and black-box settings.
Lastly, we performed adversarial shrinking and enlarging of segmentations of different anatomical structures.

All attacks were performed using PGD with 100 iterations and $\alpha = \epsilon / 20$.
We also experimented with FGSM and it did not succeed in the targeted attacks case (the predictions on adversarial examples did not resemble the target shapes and their overlap was low); we did not report these results.
For the first and the second set of experiments, we used $\epsilon \in \{0.01, 0.02, 0.04, 0.08, 0.16\}$.
For the third set of experiments, we used a fixed noise level $\epsilon = 0.04$, since it was the smallest noise level that achieved excellent performance in the heart-shaping attack experiment.
The attacks were performed on images rescaled in $[-1, 1]$ (e.g. $\epsilon = 0.01$ corresponds to 0.5\% of the image intensity range.).

The loss function $J$ (see Equation \ref{eq:pgd}) used in the attacks was the negative intersection over the union (IoU) between the predicted segmentation and the target adversarial segmentations averaged across the five structures. This was also the loss used to train the target and surrogate networks, although in that later case, the predictions were compared to the ground truth segmentations. 

\begin{figure*}[!h]
\centering
\includegraphics[width=17cm]{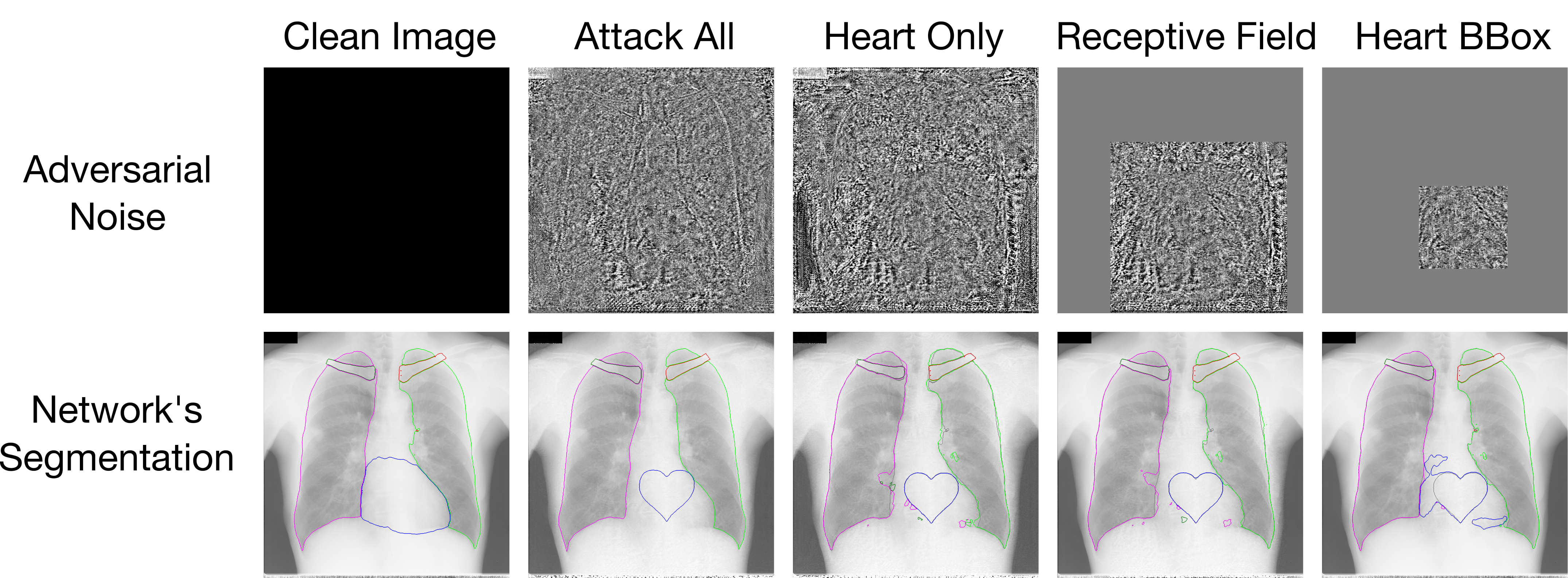}
\caption{\textbf{Influence of the type of adversarial noise on the effectiveness of targeted white-box attacks.} The types of adversarial noise added to the original image are displayed in the first row. In the second row, we show the adversarial images and the resulting network's segmentations. The heart segmentation contours are shown in blue. The light gray contour is the target segmentation of the attack. The other colors represent other structures. The first column (``Clean image'') is the original network's predictions, without any adversarial attack. The attack in the second column (``Attack All'') is optimized to make both the network's prediction fit the heart symbol and the predictions of the other structures fit the ground truths. This is the standard attack used in the rest of the article. The attack in the third column (``Heart Only'') is optimized to only make the network's prediction fit the heart symbol, without considering the predictions of the other structures. The two last columns also used this "Heart Only" setting. In the fourth column (``Receptive Field''), the adversarial noise was restricted to the network's receptive field of the ground truth heart pixels. Finally, in the last column (``Heart BBox''), the adversarial noise was restricted to a bounding box around the ground truth of the heart. The attacks shown in the figure were performed with $\epsilon = 0.04$.}
\label{fig:boxnoise}
\end{figure*}

To create the target adversarial segmentation for the structure of interest, we replaced the ground truth segmentation with the target segmentation placed at the center of mass of the ground truth segmentation. 
Although in all our attacks we only attempted to adversarially modify the segmentation of a single structure at a time, we also used target adversarial segmentations for the other structures: optimizing IoU of only the target class worsened segmentations of other structures (Figure \ref{fig:boxnoise}).
To create the target adversarial segmentation of the other structures in the image, we simply used the ground truth segmentation.
We also tried using network's predictions on non-adversarial versions of the images instead of the ground truth as the basis for target segmentations for the heart-shaping attack: this corresponds to the attack attempting to modify the heart segmentation, while keeping the rest of the segmentations exactly same. The results did not substantially differ from using the ground truth for the target segmentations quantitatively or qualitatively.

The target and surrogate networks had the same architecture and were trained using the exact same procedure. The only difference was that the training and validation sets were different, albeit they were sampled from the same set of images (i.e. the union of training and validation sets was exactly the same for both networks), resulting in highly overlapping training sets (they shared 77 out of 100 images; validation sets did not overlap).

All metrics reported in this paper, unless stated otherwise, are IoUs averaged over the test set images.

\subsection{Untargeted attacks}

\begin{figure*}[!t]
\centering
\includegraphics[width=0.75\textwidth]{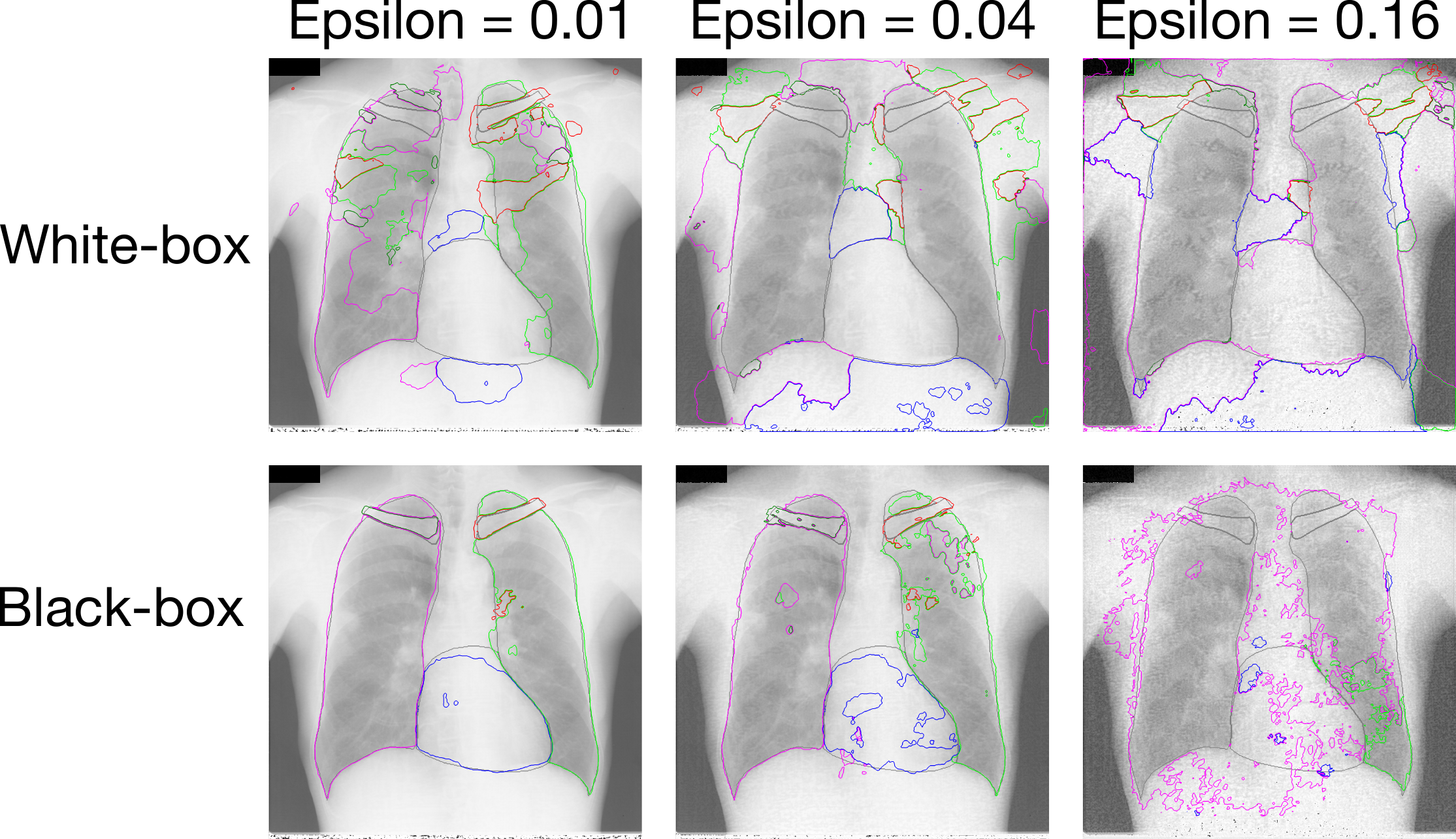}
\caption{\textbf{Untargeted attacks, minimizing the overlap between ground truth and predicted segmentations for all structures.}
This figure shows the overlay of adversarial examples and the contours of segmentations predicted by the network. The images in the first row are white box attacks, and those in the second row are black-box attacks. The columns correspond to attacks with varying noise level $\epsilon$. Blue is the contour that delineates the predicted segmentation of the heart; light green and purple, the lungs; and dark green and red, the clavicles.
Grey contours are ground truth segmentations for all structures.
Increased $\epsilon$ makes the segmentations substantially less accurate.} 
\label{fig:minimizeOverlap}
\end{figure*}

Figure \ref{fig:minimizeOverlap} shows the network's predictions for white-box and black-box adversarial examples designed to minimize the overlap between the ground truth and the predictions for different magnitudes of the noise. Table \ref{table:minimizeOverlap} shows the quantitative results on the test set. Adding almost imperceptible noise substantially deteriorated the quality of the segmentations in the white-box attack setting.
In the black-box attack setting, the performance decrease resulting from the attacks was small.

\begin{table}[!h]
\setlength{\tabcolsep}{5pt}
\caption{\textbf{White-box and black-box untargeted attacks with varying noise levels (columns).} The values in the table are IoU * 100 between the ground truth and predicted segmentations averaged over the five structures. Mean IoUs computed for non-adversarial images were 90.3 and 90.3 for the target and surrogate networks, respectively.}
\begin{center}
\begin{tabular}{c|c|c|c|c|c}
Attack type & 0.01 & 0.02 & 0.04 & 0.08 & 0.16 \\ \hline
 white-box & 30.5 &   0.8 &   0.1 &   0.0 &   0.0 \\
 black-box & 87.4 &  84.2 &  74.7 &  45.5 &  12.7 \\  
\end{tabular}

\end{center}
\label{table:minimizeOverlap}
\end{table}

\subsection{Heart-reshaping attacks}

Figure \ref{fig:boxnoise} shows adversarial heart attack performed with different settings. Since the heart attack that optimized only the heart class also disrupted the segmentations of the other surrounding structures, we used the heart attack version that optimizes the predicted segmentations of the other structures to be close to the ground truth, in addition to reshaping the heart. We used the same strategy for the resizing attacks.

Interestingly, the adversarial noise of the attack only optimizing the heart segmentation was not localized around the heart. We think this is due to the attack optimizing for achieving even lower confidence levels for the heart class for negative pixels in the entire image. The segmentations of the network were not changed by restricting the noise to the region around the heart, accounting for the receptive field. However, if the noise was restricted to the bounding box containing the heart segmentation, the heart-symbol shaping was disrupted, indicating that the attack modifies neighbourhoods of pixels to change their classification.

\begin{figure*}[!t]
\centering
\includegraphics[width=\textwidth]{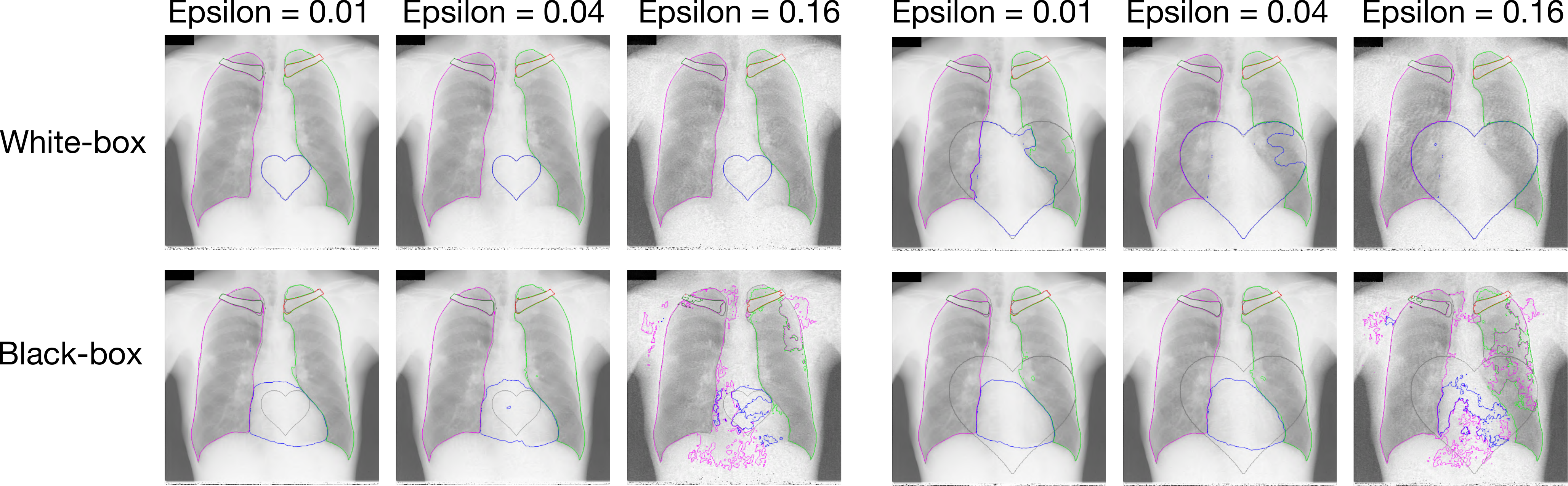}
\caption{\textbf{Targeted heart reshaping attacks.} The light gray contour is the target segmentation of the attack, which intended to reshape the heart segmentation as a large heart symbol and keep the segmentations of other structures close to the ground truth. The blue contour delineates the heart segmentation predicted by the network. The targeted heart shapes are small on the right and large on the left. All attacks were computed using PGD with 100 iterations. The images in the first row are white box attacks, and those in the second row are black-box attacks. Each column corresponds to a different noise level $\epsilon$.}
\label{fig:BigHeartShape}
\end{figure*}

Figure \ref{fig:BigHeartShape} shows white-box and black-box heart attack results with small and large heart symbols as targets.
Tables \ref{table:heartAttackIoU} and \ref{table:heartAttackIoUOtherStruct} show the quantitative results.
With a sufficient level of adversarial noise, white-box attacks achieved very high overlap between the target and predicted segmentations for the heart class, as well as for the other structures (note that for the other structures the target was the ground truth).
The attack using the larger heart shape as the target appeared more challenging, with higher $\epsilon$ needed to achieve very high overlap.
In some cases, as shown in Figure \ref{fig:BigHeartShape}, the part of the heart symbol extending into lungs was segmented as background instead of heart class.
This, however, was reduced with higher $\epsilon$.

With $\epsilon = 0.04$, for both small and large heart attacks, the overlap of predictions with the target heart symbol was larger than the overlap of predictions on non-adversarial images with the original ground truth (87.3 IoU).
Furthermore, this level of noise appeared quite visually subtle.
We thus used this $\epsilon$ for our last set of experiments.

Black-box heart-shaping attacks were completely unsuccessful for all noise levels.
The predicted heart segmentations generally did not resemble a heart symbol and had low overlap with the target segmentation.
At noise levels of $\epsilon \leq 0.04$, black-box attacks did not seem to affect the segmentations significantly: the overlaps of the predictions with the ground truth segmentations for heart (see the lower part of Table \ref{table:heartAttackIoU}) and other structures (see Table \ref{table:heartAttackIoUOtherStruct}) was close to that for non-adversarial images.
With the highest noise level $\epsilon = 0.16$, black-box attacks substantially decreased the quality of segmentations for all structures.

%When using black box attacks, the segmentation of the heart was deteriorated but did not resemble a heart symbol.

\begin{table*}[]
\caption{\textbf{White-box and black-box targeted heart attacks with varying noise level.}
The values in the upper part of the table are intersection over union (IoU) $\times$ 100 computed between predicted and \textit{adversarial target} heart segmentations.
The values in the lower part of the table are IoU $\times$ 100 between predicted and original \textit{ground truth} heart segmentations.
IoUs for non-adversarial images computed between predicted and ground truth segmentations for the heart class were 87.3 and 87.5 for the target and surrogate networks, respectively.}
\begin{center}
\begin{tabular}{c|c|c|clclclclcl}
\multicolumn{1}{l|}{\multirow{2}{*}{}}                                                     & \multirow{2}{*}{Heart symbol size} & \multirow{2}{*}{Attack type} & \multicolumn{10}{c}{Noise level ($\epsilon$)}                                                                                           \\ \cline{4-13} 
\multicolumn{1}{l|}{}                                                                      &                                    &                              & \multicolumn{2}{c}{0.01} & \multicolumn{2}{c}{0.02} & \multicolumn{2}{c}{0.04} & \multicolumn{2}{c}{0.08} & \multicolumn{2}{c}{0.16} \\ \hline
\multirow{4}{*}{\begin{tabular}[c]{@{}c@{}}Predictions\\ vs.\\ Attack Target\end{tabular}} & \multirow{2}{*}{small}             & white-box                    & \multicolumn{2}{c}{88.0} & \multicolumn{2}{c}{96.9} & \multicolumn{2}{c}{97.8} & \multicolumn{2}{c}{98.4} & \multicolumn{2}{c}{98.6} \\
                                                                                           &                                    & black-box                    & \multicolumn{2}{c}{34.3} & \multicolumn{2}{c}{35.3} & \multicolumn{2}{c}{36.8} & \multicolumn{2}{c}{41.2} & \multicolumn{2}{c}{28.9} \\ \cline{2-13} 
                                                                                           & \multirow{2}{*}{large}             & white-box                    & \multicolumn{2}{c}{66.7} & \multicolumn{2}{c}{83.2} & \multicolumn{2}{c}{95.2} & \multicolumn{2}{c}{98.7} & \multicolumn{2}{c}{99.1} \\
                                                                                           &                                    & black-box                    & \multicolumn{2}{c}{43.3} & \multicolumn{2}{c}{43.2} & \multicolumn{2}{c}{43.4} & \multicolumn{2}{c}{38.9} & \multicolumn{2}{c}{13.1} \\ \hline
\multirow{4}{*}{\begin{tabular}[c]{@{}c@{}}Predictions\\ vs.\\ Ground Truth\end{tabular}}      & \multirow{2}{*}{small}             & white-box                    & \multicolumn{2}{c}{38.4} & \multicolumn{2}{c}{34.8} & \multicolumn{2}{c}{34.4} & \multicolumn{2}{c}{34.6} & \multicolumn{2}{c}{34.4} \\
                                                                                           &                                    & black-box                    & \multicolumn{2}{c}{85.5} & \multicolumn{2}{c}{85.1} & \multicolumn{2}{c}{84.0} & \multicolumn{2}{c}{73.4} & \multicolumn{2}{c}{22.0} \\ \cline{2-13} 
                                                                                           & \multirow{2}{*}{large}             & white-box                    & \multicolumn{2}{c}{62.5} & \multicolumn{2}{c}{50.1} & \multicolumn{2}{c}{43.7} & \multicolumn{2}{c}{42.0} & \multicolumn{2}{c}{41.8} \\
                                                                                           &                                    & black-box                    & \multicolumn{2}{c}{85.8} & \multicolumn{2}{c}{85.2} & \multicolumn{2}{c}{84.8} & \multicolumn{2}{c}{77.8} & \multicolumn{2}{c}{26.0}
\end{tabular}
\end{center}
\label{table:heartAttackIoU}
\end{table*}

\begin{table}[]
\caption{\textbf{White-box and black-box targeted heart attacks with varying noise level, performance on structures other than the heart.}
In order to only affect the heart when using adversarial heart attacks and avoid disrupting segmentations of other structures, we set the adversarial target for the other structures to be the ground truth segmentation.
In the table, we show intersection over union (IoU) $\times$ 100 computed between predicted and adversarial target (same as the ground truth) segmentations averaged all structures but the heart.
This IoU for non-adversarial images was 91.0 for the target network and 90.9 for the surrogate network.
}
\begin{center}
\begin{tabular}{c|c|clclclclcl}
\multirow{2}{*}{\begin{tabular}[c]{@{}c@{}}Heart \\ size\end{tabular}} & \multirow{2}{*}{Attack type} & \multicolumn{10}{c}{Noise level ($\epsilon$)}                                                                                            \\ \cline{3-12} 
                                                                       &                              & \multicolumn{2}{c}{0.01}  & \multicolumn{2}{c}{0.02} & \multicolumn{2}{c}{0.04} & \multicolumn{2}{c}{0.08} & \multicolumn{2}{c}{0.16} \\ \hline
\multirow{2}{*}{small}                                                 & white-box                    & \multicolumn{2}{c}{98.9}  & \multicolumn{2}{c}{99.5} & \multicolumn{2}{c}{99.7} & \multicolumn{2}{c}{99.8} & \multicolumn{2}{c}{99.8} \\
                                                                       & black-box                    & \multicolumn{2}{c}{92.9} & \multicolumn{2}{c}{93.6} & \multicolumn{2}{c}{94.5} & \multicolumn{2}{c}{93.1} & \multicolumn{2}{c}{74.6} \\ \hline
\multirow{2}{*}{large}                                                 & white-box                    & \multicolumn{2}{c}{86.7}  & \multicolumn{2}{c}{95.2} & \multicolumn{2}{c}{99.4} & \multicolumn{2}{c}{99.6} & \multicolumn{2}{c}{99.6} \\
                                                                       & black-box                    & \multicolumn{2}{c}{78.4}  & \multicolumn{2}{c}{79.1} & \multicolumn{2}{c}{80}   & \multicolumn{2}{c}{79.5} & \multicolumn{2}{c}{64.5}
\end{tabular}
\end{center}
\label{table:heartAttackIoUOtherStruct}
\end{table}

\subsection{Resizing attacks}

\begin{table*}[h]
\setlength{\tabcolsep}{5pt}
\caption{\textbf{Targeted attacks resizing segmentations of different anatomical structures, quantitative results.} The values in the table are IoU * 100 between predicted and adversarial target segmentations with different resizing coefficients for the target class: the heart, clavicles, or lungs.
The first column is the IoU computed between predictions and the ground truth for clean images.
The mean IoU * 100 with respect to other structures than the target was 98.9 or higher in all cases.
}
\begin{center}
%\begin{tabular}{c|c|c|c|c|c}
%Resizing factor & 1.0, no attack & 0.6  & 0.8  & 1.2  & %1.5  \\ \hline
%Heart         & 87.3 & 98.4 & 98.6 & 98.8 & 92.4 \\
%Clavicle      & 86.8 & 92.7 & 97.1 & 98.5 & 91.5 \\
%Lung          & 95.2 & 72.0 & 87.0 & 88.9 & 74.1
%\end{tabular}
\begin{tabular}{c|c|c|c|c|c|c|c|c}
Structure      & 1.0, no attack & 0.4  & 0.6  & 0.8  & 1.2  & 1.4  & 1.6  & 1.8  \\
\hline
heart          & 87.3           & 97.5 & 98.3 & 98.8 & 99.4 & 99.1 & 97.1 & 95.1 \\
\hline
left clavicle  & 86.6           & 90.8 & 93.8 & 98.6 & 98.9 & 98   & 95.1 & 88.8 \\
right clavicle & 87             & 92.2 & 95.4 & 98.8 & 99.1 & 97.7 & 94.5 & 88.1 \\
\hline
left lung      & 95             & 91.9 &   93.0 &   95.7 &   95.4 &   89.9 &   83.5 &   78.3 \\
right lung     & 95.4           & 95.5 &   97.1 &   97.9 &   96.3 &   91.7 &   87.1 &   83.4 \\
\end{tabular}
\end{center}
\label{table:size}
\end{table*}

\begin{figure*}[!h]
\centering
\includegraphics[width=12cm]{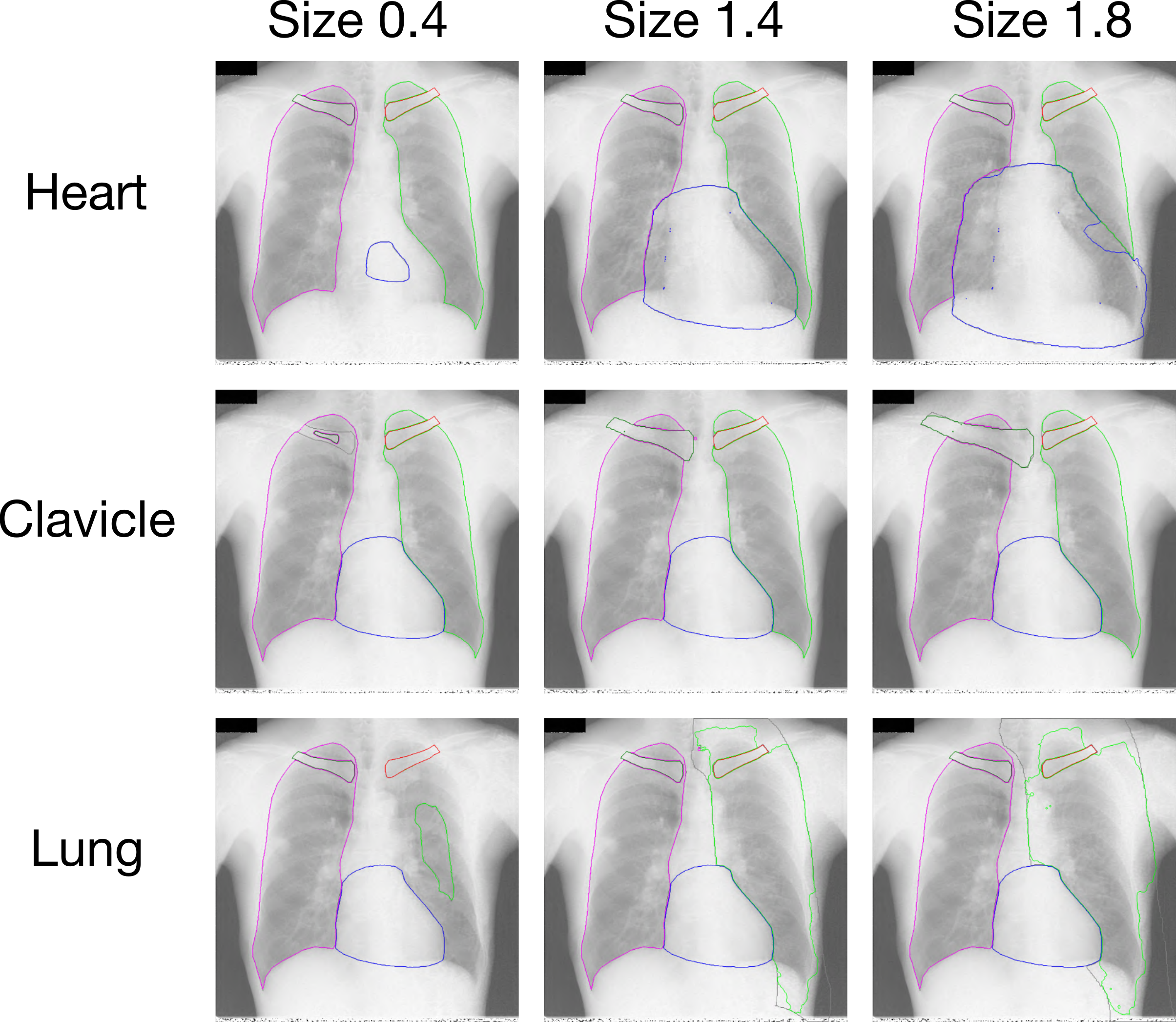}
\caption{\textbf{Targeted white-box adversarial attacks to resize segmentations.} Results are shown for the heart (blue), left clavicle (dark green), and right lung (light green) separately and for different resizing factors (Size). The light gray contour is the target segmentation of the attack, which intends to modify the size of the segmentation of the target structure and keep the segmentations of other structures close to the ground truth.}
\label{fig:size}
\end{figure*}

Figure \ref{fig:size} and Table \ref{table:size} show that adversarial attacks could resize the segmentations predicted by the network.
Clavicle and heart resizing generally performed well, also in cases of more extreme resizing.
Reducing the size of clavicles, however, often caused a change in the lung segmentation on the side of the target clavicle, making the lung close around it, despite the region around it was labeled as background in the target segmentation.
An example of that can be seen in Figure \ref{fig:size}.
This happened at least to some degree in most of the testing images.
Heart segmentations with enlarging attacks with higher resizing coefficients (1.4 and larger) in some cases failed to extend into lung regions to fit the target segmentations.
Resizing attacks seemed to be less effective for the lungs compared to the other two organs.
Lung segmentation predictions on adversarial images overlapped less with the resized segmentation targets than predictions on non-adversarial images overlapped with the ground truth with more extreme resizing, for both shrinking and enlarging attacks.
Enlarging attacks with high resizing coefficients were particularly problematic: predicted segmentations often did not extend into background regions in the lower corners of the images corresponding to air, or the upper image regions (around neck and above shoulders), or the region between the lungs.
In the case of extreme lung shrinking attacks, the network would still segment as the target lung small regions around the clavicle on the same side with the lung.
This may have a common cause with clavicle shrinking attacks leading to extending lung segmentations so that they surround the target clavicle.

\section{Discussion}

White-box attacks, both untargeted and targeted, were generally quite successful in our study: predicted network segmentations could be effectively disrupted or accurately reshaped in a targeted way, even into segmentations with anatomically implausible shape or location, using visually subtle adversarial noise only.

However, our study identified some limits for targeted white-box attacks.
Firstly, it was difficult to reshape segmentations so that they extend into regions with texture and intensity uncharacteristic of the target organ, particularly regions further away from the organ's boundaries.
This was supported by several observations: larger noise levels $\epsilon$ were needed for heart attacks with large heart symbol targets (see Figure \ref{fig:BigHeartShape}) to extend into lung regions. A similar effect was observed with extreme heart enlargement attacks; lung segmentations with lung enlargement attacks often did not extend into background regions with higher or lower intensity than lungs.
Generally, less extreme changes (less resizing and less extending into other structure's regions) were easier to achieve.
Secondly, adversarially changing segmentations in ways that conflict with relationships between classes learned by the network can be challenging.
In our study, this is exemplified by the network trying to preserve adjacency of segmentations of clavicles and lungs (from the same side) despite the adversarial attack targeting a segmentation where these structures a not adjacent (see lung and clavicle shrinking experiments in Figure \ref{fig:size}).
It would also be reasonable to expect lower attack effectiveness with adversarial attacks that contradict other learnt priors such as shapes of organs. White-box targeted heart symbol attacks did not seem to encounter difficulties in overwriting the anatomical shape of the heart. This suggests that the network did not really learn this anatomical shape.

Interestingly, targeted heart attack did not require larger noise level $\epsilon$ compared to the untargeted attack to have a very high performance: at $\epsilon = 0.02$, heart attacks using both large and small heart symbol sizes already fitted to the target segmentations of the attack, and for the same noise level $\epsilon$ the untargeted attack achieved almost zero IoU with the ground truth.
However, our resizing experiments showed that, even with $\epsilon = 0.04$, some targeted segmentation changes did not succeed most of the time. For example, making the network segment regions located outside the body--in the corners of the image--failed most of the time.
This suggests that targeted attacks optimizing for large changes could require more noise compared to untargeted attacks.

Black-box attacks were substantially less successful.
Even in the untargeted attack case, the performance of the network was only slightly reduced (Table \ref{table:minimizeOverlap} and Figure \ref{fig:minimizeOverlap}).
Some other studies that performed black-box adversarial attacks on medical image analysis networks reported much higher performance deterioration with untargeted attacks, including PGD attacks \cite{finlayson2018adversarial,wetstein2020adversarial}.
Low performance of black-box attacks in our study could be due to our target and surrogate networks having different training and validation data: differences in training and validation sets were previously reported to reduce performance of black-box attacks \cite{szegedy2013intriguing,wetstein2020adversarial}.

In the targeted case case, black-box attacks were unsuccessful. the predicted heart segmentations did not resemble heart symbol shape and had low overlap with the target adversarial segmentation (Table \ref{table:heartAttackIoU} and Figure \ref{fig:BigHeartShape}).
We believe using higher noise level $\epsilon$ is unlikely to improve performance, since using the highest $\epsilon = 0.16$ resulted in deterioration of predicted segmentation for all structures both with respect to the ground truth and targeted segmentations (Tables \ref{table:heartAttackIoU} and \ref{table:heartAttackIoUOtherStruct}). 
This is not an unsurprising result, since targeted black-box attacks were already reported to be challenging for image classification \cite{xie2019improving}. For example, in the study of Xie et al. \cite{xie2019improving}, regular targeted PGD achieved only ~10-20\% success rate in attacking networks trained to classify ImageNet (see Table 1 from the study). Black-box attacks on segmentation networks had until now not been thoroughly studied.
Using more advanced algorithms, such as data-augmentation-based attacks \cite{xie2019improving}, could potentially increase the success of black-box segmentation-shaping attacks.

Overall, our results suggest that the success of targeted adversarial attacks is contingent on the attacks being white-box. We suspect these findings could be generalized to other studies which found white-box adversarial segmentation attacks to be successful \cite{xie2017,ozbulak2019}. Switching to black-box attacks in those datasets may also substantially decrease the effectiveness of the attacks.

A limitation of this study is that we only assessed visual perceptibility of adversarial noise under the default setting of the window level and width.
In settings where careful visual inspection of the adversarial example is possible, the noise may be more visible to observers inspecting images under different brightness and contrast settings. 
This makes it difficult to ensure that studied adversarial attacks cannot be discovered by human observers.
More subtle attacks than those presented in this article may be employed. For example, Kugler et al. \cite{kugler2018} and Su et al. \cite{su2019} have proposed to only slightly modify the image by performing one pixel attacks. Adversarial attacks can also be performed as a careful rotation of the image \cite{engstrom2019}. 

To counter adversarial examples, researchers have built models that are robust to adversarial noise \cite{najafi2019}, or models that can identify adversarial samples \cite{ma2020}. Akhtar et al. \cite{akhtar2018}, Yuan et al. \cite{yuan2019} and Biggio et al. \cite{biggio2018} reviewed possible defenses and proposed taxonomies to classify defenses. 
These works focused mostly on classification networks but similar approaches could be applied to reduce the susceptibility of segmentation networks to adversarial attacks.
However, our results suggest that segmentation networks cannot easily be attacked in a targeted manner when the model weights are unknown. In situations where the exact model weights cannot be known to possible attackers, the risk of adversarial attacks manipulating segmentation outcomes appears limited.

\section{Conclusion}
In this work, we showed that it is possible to manipulate images, imperceptably, such that state-of-the-art medical image segmentation networks provide a specific and wrong segmentation. However, such targeted attacks were only successful in the white-box setting in which the attacker knows all details of the network including its architecture and weights. In a setting in which attacks were computed based on a model with the same architecture, but without knowing the weights, targeted attacks were no longer successful. This suggests that there is limited risk with targeted adversarial attacks manipulating segmentation.

\section{Acknowledgments}
This research was funded by The Netherlands Organisation for Health Research and Development (ZonMw) Project 104003005, with additional support of Netherlands Organisation for Scientific Research (NWO), project NWO-EW VIDI 639.022.010 and project NWO-TTW Perspectief Programme P15-26. This research is part of the research project Deep Learning for Medical Image Analysis (DLMedIA) with project number P15-26, funded by the Netherlands Organisation for Scientific Research (NWO). The computations were carried out on the Dutch national e-infrastructure with the support of SURF Cooperative.

{\small
\bibliographystyle{ieee}
\bibliography{egbib}
}

\end{document}